\documentclass[12pt]{article}
\pdfoutput=1
\usepackage[letterpaper, margin=1in]{geometry}
\usepackage{amsmath, amssymb}
\usepackage{cite}
\usepackage{graphicx}
\usepackage[font=small,labelfont=bf]{caption}
\usepackage[normalem]{ulem}
\usepackage{makecell}
\usepackage{mathabx} 

\usepackage{slashed}
\usepackage{cancel}
\usepackage{bm}
\usepackage{verbatim}
\usepackage{tabularx}
\allowdisplaybreaks

\usepackage[utf8]{inputenc}
\usepackage{graphicx}
\usepackage{caption}
\usepackage{subfigure}
\usepackage{hhline}
\usepackage{amsmath}
\usepackage{amsfonts}
\usepackage{amssymb}
\usepackage[dvipsnames]{xcolor}
\RequirePackage{luatex85}
\usepackage{braket}
\usepackage{array} 
\usepackage{pict2e}
\usepackage{multirow}
\usepackage[compat=1.1.0]{tikz-feynman}
\usepackage[super]{nth}
\usepackage{float}
\pagecolor{white}
\definecolor{babyblueeyes}{rgb}{0.63, 0.79, 0.95}
\definecolor{carminepink}{rgb}{0.92, 0.3, 0.26}

\usepackage{fancyhdr}

\makeatletter
\newcommand{\preprint}[1]{\gdef\@preprint{#1}}
\newcommand{\@preprint}{} 

\fancypagestyle{firstpage}{%
  \fancyhf{}
  \fancyhead[R]{\@preprint}
}
\makeatother

\usepackage[most]{tcolorbox}

\newcommand{\Eq}[1]{Eq.~(\ref{eq:#1})}
\def\[{\begin{equation}}
\def\]{\end{equation}}

\usepackage{stackengine}
\stackMath

\definecolor{DeepPurple}{rgb}{0.294, 0.180, 0.514}
\definecolor{HypotheticBlue}{rgb}{0.216, 0.494, 0.722}
\definecolor{HypotheticGreen}{rgb}{0.302, 0.686, 0.290}
\definecolor{HypotheticOrange}{rgb}{1.000, 0.498, 0.000}

\definecolor{darkgreen}{rgb}{0,0.6,0}
\definecolor{darkorange}{rgb}{0.99,0.5,0}
\definecolor{turquoise}{rgb}{0.0,0.7,0.7}

\definecolor{amethyst}{rgb}{0.6, 0.4, 0.8}
\definecolor{antiquefuchsia}{rgb}{0.57, 0.36, 0.51}
\definecolor{blue-violet}{rgb}{0.54, 0.17, 0.89}

\definecolor{customred}{rgb}{0.9, 0.36, 0.36}
\definecolor{customyellow}{rgb}{.92, 0.79, 0.0}
\definecolor{customorange}{rgb}{0.95, 0.55, 0.00}
\definecolor{customgreen}{rgb}{0.17, 0.67, 0.11}
\definecolor{custompurple}{rgb}{0.41, 0.16, 0.68}

\definecolor{custombrown}{rgb}{0.45, 0.25, 0.1}

\definecolor{DarkBlue}{rgb}{0,0,0.6}
\definecolor{MildBlue}{rgb}{0.5,0.5,1}

\usepackage{setspace}
\usepackage{authblk}
\usepackage{physics}
\usepackage{multirow}

\usepackage{bbm}

\usepackage{dsfont}

\usepackage{etoolbox}

\let\OLDthebibliography\thebibliography
\renewcommand\thebibliography[1]{
	\OLDthebibliography{#1}
	\setlength{\parskip}{3.0pt plus 2.5pt minus 1.0pt}
	\setlength{\itemsep}{3.0pt plus 2.5pt minus 1.0pt}
}

\usepackage[colorlinks=true,linktocpage=true,linkcolor=darkorange,citecolor=customgreen,urlcolor=customgreen]{hyperref}

\definecolor{goodgreen}{rgb}{0,.6,0.4}

\newcommand{\Q}{\mathbf{Q}}
\newcommand{\Tc}{T_\text{c}}
\newcommand{\Td}{T_\text{dark}}
\newcommand{\Trh}{T_\text{RH}}
\newcommand{\dNeff}{\Delta N_\mathrm{eff}}

\newcommand{\alphad}{\alpha_\text{dark}}
\newcommand{\darkrho}{\rho_\text{dark}}
\newcommand{\dotdarkrho}{\dot{\rho}_\text{dark}}

\newcommand{\trh}{T_{\mathrm{RH}}}

\usepackage{cleveref}
\crefname{table}{Tab.}{Tabs.}
\crefname{equation}{Eq.}{Eqs.}
\crefname{eqnarray}{Eq.}{Eqs.}
\crefname{appendix}{Appendix.}{Apps.}
\crefname{section}{Sec.}{Secs.}
\crefname{figure}{Fig.}{Figs.}

\title{Quirks Live in Cool Universes}

\author{Pouya Asadi$^{1}$, Graham D. Kribs$^{2,3}$, Markus A. Luty$^4$ \\
{\small \color{custompurple} 
\texttt{pasadi@ucsc.edu, kribs@uoregon.edu, maluty@ucdavis.edu},}
\\\vskip 1em
{\small \textit{${}^1$University of California Santa Cruz and Santa Cruz Institute for Particle Physics,}} \\
{\small \textit{Santa Cruz, CA 95064, U.S.A}} \\\vskip1em
{\small \textit{${}^2$Institute for Fundamental Science and Department of Physics,}}\\
{\small \textit{University of Oregon, Eugene, OR 97403, USA}} \\\vskip1em
{\small \textit{${}^3$Theoretical Physics Department, CERN, 1211 Geneva 23, Switzerland}} \\\vskip1em
{\small \textit{${}^4$Center for Quantum Mathematics and Physics (QMAP),}}\\
{\small \textit{University of California, Davis, CA 95616, USA}}
}

\date{}

\preprint{CERN-TH-2025-243}

\begin{document}

\maketitle
\thispagestyle{firstpage} 

\vspace{-0.3in}

\begin{abstract}
We demonstrate that cosmological observations
place strong 
bounds on the 
reheat temperature $T_\text{RH}$ of the Standard Model (SM) 
in minimal models of `quirks' -- heavy fermions transforming under the SM 
gauge group together with 
a new non-Abelian
gauge interaction with a confinement scale far below the mass of the fermions. 
These models have unique collider signals associated with the confining flux strings, which cannot break due to the large mass of the quirks.
Our work shows that in these models $T_\text{RH} \lesssim \mathcal{O}(100)$~GeV for the entire `quirky' parameter space where the effects of the flux string are important.
These bounds are in tension with most models of baryogenesis, showing that the discovery of quirks at colliders can have far-reaching implications for cosmology.
The bounds arise because the 
irreducible relic abundance 
of glueballs from UV freeze-in, 
combined with their long lifetimes, leads to constraints from the disruption of BBN, distortions of the CMB, excess $\gamma$-rays, an over-abundance of self-interacting dark matter, and contributions to $\Delta N_{\rm eff}$.
The glueball freeze-in abundance has a strong dependence on
$T_\text{RH}$, 
making the bounds relatively insensitive to strong interaction uncertainties. 
The bounds are robust to the SM quantum numbers of the quirks and the presence of Yukawa couplings with the Higgs. 
In non-minimal extensions of the model where the glueballs can decay to an additional dark sector, the bounds remain for models where the flux string has a macroscopic length at colliders.
We also show 
that for quirk masses above $\sim 10$~TeV, the dark glueballs can be the dominant component of dark matter.
This work illustrates a striking 
connection between quirky collider signals and cosmological probes of new physics, strengthening the case for targeted quirk searches at colliders.

\end{abstract}

\newpage 

\begin{spacing}{1.2}

\tableofcontents

\section{Introduction}
\label{sec:intro}

The Standard Model (SM) of particle physics accurately describes all the interactions
of elementary particles that have been experimentally probed in the laboratory so far,
but it leaves many important problems unanswered.
It does not account for dark matter, the matter-antimatter
asymmetry, or cosmic inflation.
Furthermore, it implies that the electroweak symmetry breaking scale is sensitive
to ultraviolet (UV) scales such as the Planck scale, resulting in the so-called hierarchy problem.
A wide range of models for physics beyond the SM
has been proposed to address these and other open questions in the SM\@.
The mixed success of these models, together with the wide range of problems in the SM that need to be addressed, suggests that the new physics that addresses these questions is unlikely to be simple or minimal.
This is largely due to the fact that the absence of signals for physics beyond the Standard Model (BSM),
combined with the precise agreement of experiments with the SM 
predictions (notably flavor physics and 
precision electroweak tests), implies that 
any new physics beyond the SM is 
very tightly constrained.
Given this situation, it is important to explore simple extensions of the SM
that preserve its successes, even if they do not directly address 
its shortcomings.

In this paper, we study one such extension, consisting of a new sector with a SU($N$) gauge group coupled to heavy fermions.
This is a very minimal extension of the SM with two parameters,
the mass $m_\Q$ of the new fermions, and the 
strength of the gauge coupling, which can be parameterized by its confinement scale $\Lambda$.
We are interested in the regime where $m_\Q \gg \Lambda$,
and $m_\Q \gtrsim 100$~GeV\@.
If the new fermions are produced in a high-energy collision,
there will be a dark gauge flux string connecting them,
analogous to the color string connecting quarks in QCD\@.
This string can break only by nucleating a pair of heavy quarks, a process that is
analogous to Schwinger pair production of charged particles in an electric field \cite{Schwinger:1951nm}.
The probability for this process is exponentially 
suppressed by $\sim e^{-m_\Q^2/\Lambda^2}$, so if the fermions are sufficiently
heavy, the strings essentially never break \cite{Okun:1979tgr,Okun:1980mu,Kang:2008ea}.
This can give rise to exotic experimental signals, so the heavy fermions are called `quirks.' 
The quirks are charged under at least part of the SM gauge group, but the new strong gauge bosons do not interact directly with the SM, so we will refer to
the string, the glueballs made from them, and the new confinement scale as `dark.'%
\footnote{One of the authors of this paper thinks the SU($N$) gauge interactions should be called `infracolor' because of the low confinement scale, but they have been out-voted. }

Quirks can arise in a variety of BSM frameworks, such as little Higgs models \cite{Cai:2008au}, folded supersymmetry \cite{Burdman:2006tz,Burdman:2008ek}, twin Higgs scenarios \cite{Craig:2015pha,Craig:2016kue}, and confining dark sectors \cite{Kribs:2009fy}, among others.
Confining dark sectors with a sufficiently large scale hierarchy to accommodate quirks can also exhibit exotic dynamics during a first-order confinement phase transition, which can significantly deplete the abundance of quirks \cite{Asadi:2021pwo,Asadi:2021yml,Asadi:2022vkc,Gouttenoire:2023roe}.

In this paper, we focus on the case where the quirks are charged under the SM gauge interactions, making
them potentially visible in collider experiments. 
For minimal choices of the SM quantum numbers of the quirks, this 
technically natural extension of
the SM preserves all of the successes of the SM
making it a simple and plausible
model for new physics.
On the other hand, the collider signals are very exotic, since the quirks are
connected by strings that prevent them from separating.
A rough measure of the maximum length of a string produced at a collider is
obtained by assuming that the initial quirk kinetic energy $\sim m_\Q$ is converted
into string tension energy:
\[
\ell = {\rm string\ length} \sim \frac{m_\Q}{\sigma_{\mathrm{str}}^2} \sim
1~\text{cm} 
\left( \frac{m_\Q}{100~\text{GeV}} \right)
\left( \frac{\Lambda}{\text{keV}}
\right)^{-2}.
\label{eq:str_length}
\]
Here $\sigma_{\mathrm{str}}$ is the string tension and we have parameterized the strong scale by the 
deconfinment temperature $\Lambda \equiv \Tc$ using
$T_c/\sqrt{\sigma_{\mathrm{str}}} \simeq 0.6$ \cite{Boyd:1996bx,Lucini:2012wq}. 
The confinement scale $\Lambda$ can be naturally well below the
electroweak scale, so the string length can be anywhere from microscopic to 
macroscopic, resulting in a wide 
range of very exotic collider signals \cite{Kang:2008ea}.
Although there have been a number of interesting proposals for such searches
\cite{Chacko:2015fbc,Farina:2017cts,Knapen:2017kly,Evans:2018jmd,Curtin:2025ksm,Condren:2025czc}, so far only a single dedicated search for quirks has
been performed at colliders \cite{D0:2010kkd}.

In this paper, we consider the connection between quirks and cosmology.
At energies below $m_\Q$, this theory consists of the SM
plus a `dark' Yang-Mills theory coupled through suppressed (dimension-8) `portal' 
interactions. 
Below the dark confinmenet scale $\Lambda$, the lightest particles in the dark sector are the glueballs
in the Yang-Mills sector.
The lightest glueballs would be absolutely stable if we neglected
the portal interactions with the SM, 
but instead, they are long-lived because 
their decay proceeds through higher-dimension operators.

We show there are strong 
constraints on the relic abundance of the glueballs arising from
their decays to SM particles, their relic abundance and self-interaction, 
as well as gluons contribution to $\Delta N_\mathrm{eff}$.
These constraints require the glueball relic abundance
to be highly suppressed compared to the 
scenario where the quirks and dark gluons are in equilibrium 
with the SM early in the universe.
This strongly constrains cosmological histories 
where the SM sector is reheated to
temperatures of order $m_\Q$ or higher, since the SM interactions
of the quirks will efficiently equilibrate the quirks and dark
gluons. 
We are therefore led to consider scenarios where the SM is
reheated to a temperature $\Trh \ll m_\Q$.
In order to suppress the gluon abundance, we must assume that the reheating 
sector does not directly reheat the dark gluons.

However, there is an irreducible dark glueball abundance arising from out-of-equilibrium interactions between
the SM and the dark gluons (`freeze-in') 
\cite{McDonald:1993ex,Hall:2009bx,Elahi:2014fsa}.
We show that this sub-thermal abundance is strongly constrained by cosmological observations, ruling out models with $\Lambda \lesssim \text{GeV}$ and $m_\Q \lesssim 10~\text{TeV}$ unless the reheat temperature is below $\sim 100$~GeV.
This tightly constrains conventional models of baryogenesis which require larger reheat temperatures.
Remarkably, the parameter range where the model requires a low reheat temperature includes the entire range where quirks are accessible to current and planned collider experiments through
signals that are difficult or impossible to explain without the effects of the unbreakable string -- we refer to these as `quirky' signals.
This includes the entire region where the quirk strings are `macroscopic' or `mesoscopic', as well as a large part of the `microscopic' regime, see \cref{sec:basics} below for definitions of these regimes.

We also consider non-minimal extensions of the quirk model where the glueballs can decay to new light neutral particles.
These models are significantly more complicated, removing one of the main motivations for studying these models.
Nonetheless, we show that even such contrived scenarios do not eliminate the constraints in the macroscopic and mesoscopic regimes.

This shows that any observation of quirky signals at
colliders 
has striking implications for cosmology.
We hope that this work stimulates interest in performing searches for these 
signals at colliders.

\section{Review of Quirks and Glueballs}
\label{sec:basics}
Our analysis will focus on a simple quirk model with a confining SU($N$) gauge group and vector-like fermion $\Q$ (quirks) in the fundamental representation.
The quirks are in a vector-like representation of the SM gauge group SU(2)$_L$, and have vanishing hypercharge.
The Lagrangian is
\[
{\cal L} = -\frac{1}{4} G'^{\mu\nu}_a G'_{\mu\nu \, a} + i \bar{\Q} \slashed{D} \Q - m_{\Q} \bar{\Q} \Q \, ,
\]
where $G'_{\mu\nu \, a}$ is the field strength of the new SU($N$) gauge group.
Our benchmark model will have $N = 2$, but we will later discuss larger values of $N$, as well as various extensions of the minimal model.
We will argue that the constraints on simple extensions of the minimal model are very similar.
This model has 2 parameters, the quirk mass $m_{\Q}$ and the confinement scale $\Lambda$. 
We will be interested in values of $m_\Q$ such that the quirks are accessible to present or planned colliders, roughly $100~\text{GeV} \lesssim m_\Q \lesssim 3~\text{TeV}$ with $\Lambda \ll m_\Q$.

Below the mass $m_\Q$ the theory is a pure SU($N$) Yang-Mills model.
The lightest states in this theory are therefore glueballs.
The low-lying spectrum of these glueballs is well-established from lattice studies \cite{Fritzsch:1975tx,Jaffe:1975fd,Novikov:1981xi,Cornwall:1982zn,Bali:1993fb,Morningstar:1999rf,Chen:2005mg,Juknevich:2009ji,Juknevich:2009gg,Yamanaka:2019aeq,Yamanaka:2019yek}
(see Ref.~\cite{Mathieu:2008me} for a review).
The lightest of these glueballs is the $0^{++}$ glueball with a mass of $m_{0^{++}} \simeq 6 \Lambda$, where we define $\Lambda$ to be equal to the temperature of the confining phase transition \cite{Teper:1998kw,Lucini:2004my,Athenodorou:2020ani}.
The relic abundance and decays of this particle give rise to the constraints that we derive below.
There are additional glueballs in this theory that are stable in the limit where we only have the SU($N$) gauge interactions, but we will argue below that including the heavier glueballs and their decays can only strengthen our bounds.

The heavy quirks will also have a relic abundance, and the cosmology of these states is complicated by the confining string interactions at late times.
It is expected that this strongly dilutes the relic abundance of hadrons containing quirks after the confinement phase transition \cite{Kang:2006yd,Mitridate:2017oky}.
For $N \ge 3$ the confinement phase transition becomes first order, and the dark baryon abundance depleted during the phase transition by `squeeze-out' \cite{Asadi:2021pwo,Asadi:2021yml,Asadi:2022vkc,Gouttenoire:2023roe}, further suppressing their cosmological relevance.
As a result, for quirk masses accessible at colliders \cite{Asadi:2025vfr} (the focus of this work), the dark baryons do not have any impact on the cosmology today and are neglected in this study.

The signatures of quirks at colliders are very exotic \cite{Kang:2008ea}.
The quirks are pair-produced through SM interactions, for example Drell-Yan in our benchmark model.
As in QCD, the confining SU($N$) gauge interaction gives rise to a flux tube that connects the quirks.
The tension of the flux tube is set by $\Lambda$, while the kinetic energy of the quirks is set by $m_\Q$, so for $m_\Q \gg \Lambda$ the string has a small effect on the production and initial propagation of the quirks as long as they are not produced near threshold.
In QCD, the color flux tube can break due to the nucleation of light quarks, but there are no light particles charged under the new confining SU($N$) gauge interaction.
The process of SU($N$) flux tube breaking is analogous to Schwinger pair production, and is suppressed by $e^{-m_\Q^2/\Lambda^2}$.
This suppression means that the string never breaks as long as $m_\Q \gtrsim \text{few} \times \Lambda$, allowing the flux tube to stretch to a length $\ell \sim m_\Q / \Lambda^2$ before the quirks stop and turn around, see \Eq{str_length}.

The collider signatures depend on the typical length $\ell$ of the strings:
\begin{itemize}
\item 
$\ell \gtrsim 100~\text{km}$ 
(`\textit{megascopic}'): 
The string force is so small that it can be neglected on detector scales.
In this limit, quirks appear as a pair of stable particle tracks.
\item 
1~$\text{mm} \lesssim \ell \lesssim 100$~km
(`\textit{macroscopic}'):
The separation between the quirks can be resolved in the detector, 
and the string force is important
for the motion on detector scales.
The fact that the quirks interact independently with the detector ensures
that the quirk pair does not annihilate \cite{Knapen:2017kly,Evans:2018jmd}, except for cases where one
of the quirks is stopped in the detector.
In this regime, the quirk tracks cannot be
reconstructed with conventional tracking algorithms.
A number of proposals have been put forward to search for quirks in this
regime \cite{Farina:2017cts,Knapen:2017kly,Evans:2018jmd,Condren:2025czc}. 
\item 
$10^{-6}~\text{m}  \lesssim \ell \lesssim 1$~mm
(`\textit{mesoscopic}'):
The quirk separation is too small to be resolved in the detector, but the quirk
annihilation probability is sufficiently small that a quirk pair produced
at the interaction point makes it to the beam pipe with a radius of order $1~\text{cm}$.
Once the quirk pair are in matter, interactions with matter generate a large
angular momentum for the pair suppressing annihilation \cite{Kang:2008ea}.
In this limit, the quirks appear as a single stable particle track
that will be highly ionizing if the quirks are charged (even if the bound
state is neutral), and may have observable radiation.
A dedicated search for this signal for neutral ionizing tracks
was performed at D0 \cite{D0:2010kkd}, but this regime is comparatively less explored. 
\item 
$\ell \lesssim 10^{-6}~\text{m}$
(`\textit{microscopic}'):
The quirk pairs annihilate before reaching the beam pipe, 
so they appear as a resonance that can be
searched for using standard techniques, e.g. see Ref.~\cite{Chacko:2015fbc,Curtin:2025ksm}.
If the quirks carry color, the annihilation of the quirks is accompanied
by a large amount of soft radiation.
For larger values of $\ell$ in this range, the resonance may have a displaced
vertex. The dark glueballs in this regime can give rise to displaced vertex signals as well \cite{Forsyth:2025wks}.
\end{itemize}
In the macroscopic regime, the observation of a single event may
be sufficient to constitute a convincing discovery of a long-range string force.
In the mesoscopic regime, the bound states are sufficiently exotic that
it may be difficult to explain observed events with any other model.
If we think of these novel signal events
as an experimental observation of quirk strings,
then this requires the quirks to be in one of these regimes, 
which implies $\Lambda \lesssim \text{keV}$, see Eq.~\eqref{eq:str_length}.
The main result of this paper is that in 
this regime, 
the model is ruled out by cosmological constraints unless the reheat temperature is low.

\begin{figure}
    \centering
    \includegraphics[width=0.9\linewidth]{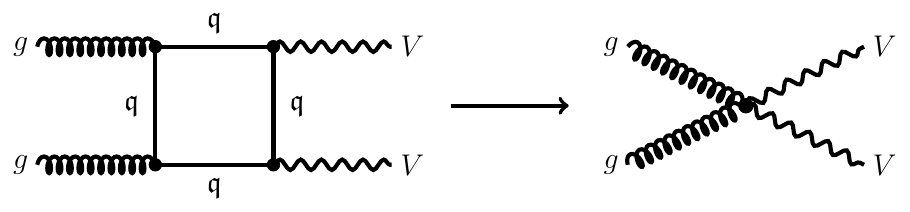}
    \caption{The diagram connecting the dark gluons to the SM in our UV theory (\textbf{left}) and the effective operator it matches onto once the heavy quarks are integrated out (\textbf{right}). 
    This work focuses on a benchmark model where the quirks are charged only under the SU(2)$_L$ of the SM, so $V$ are electroweak gauge fields.}
    \label{fig:Omatching}
\end{figure}

Integrating out the heavy quirks generates a dimension-8 interaction between the SU($N$) gauge bosons and photons, as illustrated in  Fig.~\ref{fig:Omatching}. For example
\[
\Delta {\cal L} \sim \frac{\alpha_2 \alphad}{m_\Q^4}
\tr (W^{\mu\nu} W_{\mu\nu})
\tr(G'^{\mu\nu} G'_{\mu\nu}) \, ,
\label{eq:L_IR}
\]
where $W_{\mu\nu}$ is the SU(2)$_L$ field strength, $G'_{\mu\nu}$ is the dark gluons field strength, and $\alpha_2$ and $\alphad$ are their respective gauge couplings. 
Additional dimension-8 operators with different Lorentz structure are also generated at one loop \cite{Juknevich:2009ji}, and these are relevant for the production and decay of heavier glueball states.
However, the dimension-8 operator in \Eq{L_IR} is the only one relevant for the production and decay of the $0^{++}$ glueball, and we will argue below that the heavier glueballs do not invalidate our constraints.
Later we will also consider quirks with color and/or Yukawa couplings to the Higgs.
In these models there are additional dimension-6 and dimension-8 interactions that can be important.

The glueballs interact with the SM only through higher-dimension interactions such as \Eq{L_IR}, and these are the only interactions that allow the $0^{++}$ glueball to decay.
We will be interested in low values of $\Lambda$ such that the only kinematically allowed decay of the $0^{++}$ glueball is to photons.
In this case, the decay rate of this state can be estimated as
\cite{Juknevich:2009ji}:
\begin{equation}
    \Gamma \simeq \frac{\alphad^2 (m_{0^{++}}) \alpha^2}{57600\pi}\frac{m_{0^{++}}^3 \mathbf{(F^S_{0^{++}}})^2}{ m_\Q^8},
    \label{eq:decay_width}
\end{equation}
where $m_{0^{++}}$ is the $0^{++}$ glueball mass
($m_{0^{++}} \simeq 6 \Lambda$ for $N = 2$ \cite{Teper:1998kw,Lucini:2004my,Athenodorou:2020ani}),
$\alphad (m_{0^{++}})$ is the dark gauge group structure constant evaluated at the glueball mass scale, and the $0^{++}$ glueball decay constant can be estimated as \cite{Chen:2005mg}
\begin{equation}
\mathbf{F^S_{0^{++}}} \simeq \frac{3 m_{0^{++}}^3}{4\pi \alphad (m_{0^{++}})} \, .
    \label{eq:FS0}
\end{equation}
It should be noted that Eq.~\eqref{eq:decay_width} is only a rough estimate, since it is a 1-loop calculation evaluated when $\alpha_\text{dark}$ is strong. 
Nonetheless, given the steep sensitivity to $\Lambda$, even an order of magnitude uncertainty in the lifetime amounts to negligible changes in our final results, see \cref{subsec:uncertain}.

\begin{figure}
    \centering
    \includegraphics[width=0.7\linewidth]{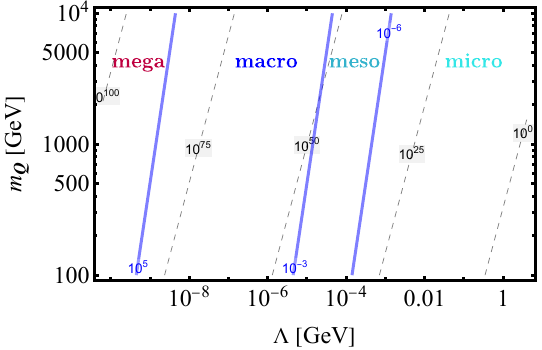}
    \caption{The lifetime (\textcolor{black}{black dashed}) of the lightest glueball in seconds as a function of the dark confinement scale and the heavy quirk mass. The quirk pair lengths in meters are denoted by \textcolor{blue}{blue} lines. These lines roughly demarcate various quirk regions 
    discussed in the bullet points in the main text.
    Note that for most of the parameter range of interest, the glueballs will be long-lived or cosmologically stable.
    }
    \label{fig:lifetime}
\end{figure}

In Fig.~\ref{fig:lifetime} we show the lifetime of the $0^{++}$ glueball as a function of $m_\Q$ and $\Lambda$, superimposed on the various regimes for collider physics discussed above. 
Note that the lifetime can span a vast range of values since it scales as $\tau_{0^{++}} \sim \Lambda^{-1} \left( \frac{m_\Q}{\Lambda} \right)^8$.
For the range of values shown in the plot, the lifetime can be very long, such that the glueballs decay after the big-bang nucleosynthesis (BBN) epoch or even remain stable throughout the entire lifespan of the universe. 
These decays are the origin of the constraints derived in this paper.
The other key ingredient is the relic abundance of the glueballs, which we turn to next.

\section{Relic Abundance of Dark Glueballs}
\label{sec:abundance}

In this section, we outline the calculation of the relic abundance of dark glueballs.
Our discussion here focuses on the physical mechanisms and order-of-magnitude estimates.
The detailed calculation of the dark gluon relic abundance used in the constraints
is discussed in Appendix \ref{app:freezein}.

As we will show below, in the parameter regime of interest to us,
there are strong cosmological constraints on the relic
dark glueballs, so the relic abundance of dark gluons
must be well below their equilibrium value in the early universe.
This can happen naturally if the reheating sector reheats the SM bath to a 
temperature $T_\text{RH}$ well below the quirk mass, assuming that the reheating
sector does not directly reheat the dark gluons.
In this case, the relic abundance of the quirks is suppressed by 
$\sim e^{-m_\Q/T_\text{RH}}$, and the gluon relic abundance is highly suppressed
because it couples to the SM through higher-dimension operators.

We assume that the reheating process is rapid, meaning that the SM fields reach the reheat temperature $T_\text{RH}$ in a time that is short compared to the inverse Hubble
parameter at reheating.
We also assume that the relic abundance of gluons vanishes at this time.
If these assumptions are not satisfied, then there will be a nonzero relic abundance 
of glueballs at the reheat temperature, and our bounds become stronger.
Here we are defining the reheat temperature as the highest temperature of the 
most recent radiation-dominated era.
For example, if the universe has a phase of matter domination due to a long-lived
particle that decays to SM particles, this can reheat the SM\@.
Any pre-existing dark gluon abundance will be diluted, but the abundance will
generally be nonzero at the reheat temperature.

With these assumptions, the dark glueballs have an \textit{irreducible} relic abundance due to out-of-equilibrium interactions with the SM
(`freeze-in') \cite{McDonald:1993ex,Hall:2009bx,Elahi:2014fsa}.
The leading interaction between the dark gluons and the SM come from dimension-8 operators, see \Eq{L_IR}. 
We focus on operators that involve the electromagnetic field strength because 
we are interested in reheat temperature below $m_W$.
The quirks in our benchmark model are not charged under SM color, so a similar coupling to the SM gluon field strength
is absent.
If it is present, it will change the relic abundance by an order-1 factor, but this does not strongly affect our results because
we are exploring a parameter space spanning many orders of magnitude.
We are also neglecting additional interactions that can arise if the quirks couple directly to the SM Higgs field.

The interaction above will populate the dark gluon sector via the process $\gamma\gamma \to g'g'$, where $g'$ is a dark gluon.
Below we estimate the relic abundance of dark glueballs due to this interaction in the freeze-in limit. 
Depending on the ratio of the kinetic energy of dark gluon $g'$ ($E_{g'} \sim \Trh$) and the dark confinement scale ($\Lambda \sim T_c$), three distinct regimes exist. We study each regime in the following subsections.

\subsection{Freeze-in with High-Energy Gluons: $\Tc/\xi \ll \Trh \ll m_\Q$}
\label{subsec:abundance_T_high}

We begin with the case where $\Trh$ is well below the quirk mass, but sufficiently
large that the dark sector comes to equilibrium with itself 
at a temperature above $\Tc$.
In this case, the dark gluons are initially produced with energy of order 
$T_\text{RH} \gg \Lambda$, so their production rate can be computed in perturbation
theory.
The gluon production rate is faster than their self-thermalization rate (since $\Trh \gg T_c$), thus we can neglect gluon self-interactions during their freeze-in.
The process is dominated at high temperature because it is mediated by an irrelevant (dimension-8)
operator, i.e.~`UV' freeze-in \cite{Elahi:2014fsa}.
As long as $\Trh \ll m_\Q$,
the dark sector will be populated well below its equilibrium abundance, 
so we can neglect the inverse reaction $g'g' \to \gamma\gamma$.

We can then estimate the relic abundance just after reheating from 
\[
\label{eq:rhodarkestimate}
\darkrho^{\text{RH}} \sim \frac{T_\text{RH}^7 \,\langle \sigma(\gamma\gamma \to g'g') \, v \rangle}{H_\text{RH}}
\sim \frac{M_\text{Pl} T_\text{RH}^{11}}{g_*(T_\text{RH})^{1/2} M^8} \, ,
\]
where $H_\text{RH}$ 
is the Hubble rate 
and $g_*(\Trh)$ is the  
number of relativistic degrees of freedom in the SM bath 
at $\Trh$.
(The precise calculation is presented in Appendix \ref{app:freezein}.) 
The energetic dark gluons then thermalize through number-changing processes
such as $g'g' \to g'g'g'$ \cite{McDonald:2015ljz}.
These processes are rapid compared to the expansion of the universe, and give an
equilibrium temperature in the dark sector
\[
\label{eq:xi}
T_\text{dark}^{\text{RH}} \equiv \xi \Trh,
\qquad
\xi \simeq \left( \frac{5\darkrho^{\text{RH}}}{\pi^2 \Trh^4}  \right)^{1/4},
\]
where the value of $\xi$ is for a SU(2) dark gauge group 
(6 bosonic degrees of freedom). 
After the reheating, the dark sector is decoupled from the SM sector, and the
transfer of energy and entropy between them is negligible.

The discussion above is valid as long as  the dark sector reheats to
a temperature above $\Tc$, {\it i.e.}~$\Trh \gg \Tc/\xi$ (see \Eq{xi}).
After reheating, we can approximate the dark sector as a gas of free relativistic gluons with temperature $T_{\rm dark}$ that starts at $T_{\rm dark}^{\rm RH}$ and decreases as the universe expands.
The gluons cool until the dark
phase transition at $\Td = \Tc$.
For our benchmark model, the dark sector gauge group is SU(2), and lattice studies 
indicate that the phase transition is second order. 
For $\Td \gg \Tc$ the dark sector is well approximated as a
gas of free gluons, and for $\Td \ll \Tc$ it can be approximated
as a gas of non-relativistic glueballs dominated by the lightest glueball with
mass $m_{0^{++}} \simeq 6 \Tc$.

The regime where $\Td \sim \Tc$ is strongly coupled, so calculations in this
regime rely on modeling or lattice data.
The microphysical time scale in this regime is set by $\Lambda$, which is
large compared to $H$.
Therefore, we expect that the evolution of the dark sector will be adiabatic,
meaning the dark sector is always close to equilibrium.
This means that we can compute the evolution of the energy density in the dark sector
if we know the \textit{equation of state} in the dark sector as a function of $\Td$.
This interpolates smoothly between the equation of state of a relativistic gas
of gluons for $\Td \gg \Tc$ and a non-relativistic gas of
glueballs for $\Td \ll \Tc$.
The evolution in this regime is complicated, and subject to significant
uncertainties:
\begin{itemize}
\item 
For $\Td \gtrsim \Tc$, the equation of state deviates significantly from that
of radiation.
This deviation is captured, for example, by the non-zero trace anomaly of the gluon plasma measured in lattice studies \cite{Boyd:1996bx}.

\item 
For $\Td \lesssim \Tc$, we expect the dark sector to be a gas of glueballs.
Because $m_{0^{++}} / \Tc \simeq 6$, these are non-relativistic.
However, they are expected to have strong self-interactions among themselves, and 
$2 \leftrightarrow 3$ interactions may lead to an era of `cannibalism' where the temperature
decreases logarithmically \cite{Carlson:1992fn, Pappadopulo:2016pkp}.
Estimates of the effect of such an era have been given in
\cite{Forestell:2016qhc,Forestell:2017wov}.

\item Lattice studies indicate that the glueball spectrum for SU(2) (SU(3)) Yang-Mills has 5 (12) absolutely stable glueball states in the limit where we turn off interactions with the SM.
All of these can decay to the SM, either through decays to lighter glueballs plus photons, or to photons \cite{Juknevich:2009gg}.
Furthermore, the $3 \leftrightarrow 2$ interactions between the glueballs dilute the abundance of glueballs with mass $m > m_{0^{++}}$ by $\sim e^{-(m - m_{0^{++}})/\Td^\text{FO}}$, where $\Td^\text{FO} < \Tc$ is the temperature in the dark sector where the heavier glueballs freeze out.
This strongly dilutes the relic abundance of the heavier glueballs \cite{Forestell:2016qhc}.
For these reasons, it is a good approximation to neglect the heavier glueballs in our constraints.

\end{itemize}

\noindent
Our benchmark SU(2) model has a second order phase transition, but
the phase transition for SU($N$) with $N > 2$ is first order
\cite{Svetitsky:1982gs,Kaczmarek:1999mm,Lucini:2003zr,Lucini:2005vg,Aoki:2006we,Saito:2011fs}.
In this case there are additional sources of uncertainty:

\begin{itemize}

\item 
The transition is expected to proceed via bubble nucleation,
and if the bubble nucleation rate is sufficiently suppressed, there can be
 supercooling during the phase transition.
We do not expect this to be significant for moderate values of $N$, because then 
the theory is governed by a single scale $\Lambda$ and has no large or 
small parameters.
\item 
A first order phase transition has a nonzero latent heat.
This is the energy (per unit volume) that must be extracted from the system in the 
deconfined phase at $\Td = \Tc$ to complete the transition.
Since the dark sector is decoupled from the SM sector, this heat can only be extracted
by the expansion of the universe.
This means that there is an era where the temperature in the dark sector remains 
constant at $\Tc$, and the dark sector
can be thought of as a mixture of the confined and
deconfined phases.
These phases can have nontrivial interactions with each other, 
and the thermodynamic properties of this form of matter are not known.
For $N = 3$, the latent heat is small ($\Delta\rho/\rho \simeq 0.08$), 
and the effect of the latent heat on the expansion of the universe can be computed
from the fact that the entropy is conserved in the dark sector:
\[
\frac{a_\text{after}}{a_\text{before}} = \left(
\frac{s_\text{before}}{s_\text{after}} \right)^{1/3}
\simeq 1.2,
\]
where we used the lattice results of \cite{Lucini:2005vg}. 

\item 
For asymptotically large values of $N$, the bubble nucleation rate is expected to be suppressed by $\sim e^{-N^2}$, since the local (intensive) thermodynamic properties of the confined and deconfined phases (such as the entropy density) differ by order $N^2$.
This suppression is verified by lattice simulations \cite{Rindlisbacher:2025dqw}, which suggest that these effects become important only for $N \gtrsim 40$ (see \cref{subsec:uncertain} below).
For arbitrarily large values of $N$, it appears that there can be an arbitrarily large amount of supercooling.
During the supercooled regime, the dark sector redshifts like radiation much longer, which tends to \emph{decrease} the final relic abundance in the dark sector, potentially weakening our bounds.
We will not attempt to incorporate these effects, so our results come with the caveat that they hold only away from large values of $N$.

\end{itemize}

\begin{figure}
    \centering
    \includegraphics[width=0.6\linewidth]{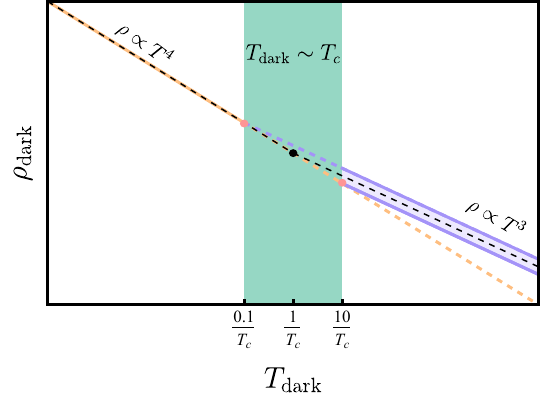}
    \caption{Schematic scaling of the energy density in the dark sector as a function of temperature.
    Our central estimate (black dashed) is given by assuming that the sector redshifts like radiation (\textcolor{customorange}{orange} lines) down to $\Td \sim \Tc$, and then redshifts like matter (\textcolor{custompurple}{purple} lines).
    The uncertainties (\textcolor{custompurple}{purple} band) are modeled by extrapolating the asymptotic equations of state through the regime $0.1 \Tc \le \Td \le 10\Tc$.
    The upper limit is obtained by assuming that the equation of state becomes non-relativistic at $\Td = 10\Tc$ (the left \textcolor{customred}{red} dot), 
    while the lower limit is obtained by assuming that the theory
    becomes non-relativistic at $\Td = 0.1\Tc$ (the right \textcolor{customred}{red} dot).}
    \label{fig:redshift}
\end{figure}

Rather than attempt to model the effects discussed above, we will estimate 
the uncertainty in the dark sector energy density today by adopting
the following simple bracketing procedure.
We assume that the dark sector can be accurately described by free gluons for
$\Td \ge 10 \, \Tc$, and free glueballs for $\Td \le 0.1 \,  \Tc$.
In the regime $0.1 \,  \Tc \le \Td \le 10 \,  \Tc$ we assume that the uncertainty can
be bracketed by extrapolating the equation of state from above and below,
as illustrated in \cref{fig:redshift}. 
This gives at most an order of magnitude uncertainty in the
final energy density.
Because the relic density is proportional to a high power of $\Trh$ 
(see Eq.~(\ref{eq:rhodarkestimate})), 
the final constraints on $\Trh$ are insensitive to such an
uncertainty in the relic abundance.

\subsection{Freeze-in with Intermediate-Energy Gluons: $T_c \ll \Trh \ll T_c/\xi$}
\label{subsec:abundance_T_mid}

In this regime, the gluons are initially produced with energy of order $\Trh \gg \Tc$,
but the subsequent self-interactions cool the gluons to a temperature
below $\Tc$.
The details of the phase transition are quite complicated and uncertain, 
as discussed in the previous subsection.
If the phase transition is second order, the timescale for the cooling process
is expected to be of order $\Lambda$, which is rapid compared to the expansion
of the universe.
In this case, the energy density of glueballs after the phase transition is given simply by equating it to the gluon energy density produced by the
initial reheating process.
What is not known is the phase space distribution of the glueballs after the phase transition.
The simplest possibility is that the glueballs are in kinetic and chemical
equilibrium right after the phase transition.
However, it is also possible that they are out of equilibrium, and come to equilibrium 
via their self-interactions.
In particular, if the gluons are not in chemical equilibrium, they will come to
equilibrium \emph{via} $3\to 2$ processes.
This results in a period of cannibalism during which the temperature of the
dark sector decreases only logarithmically.
This has the effect of delaying the completion of the phase transition, resulting in a 
\emph{larger} relic abundance for the dark glueballs.

If the phase transition is first order, the expansion of the universe
must remove the latent heat from the dark sector, and the transition cannot
be treated as instantaneous.
As discussed in the previous subsection, the latent heat is not large for 
$N = 3$, but is expected to scale as $N^2$ for large $N$ \cite{Lucini:2005vg}.
A significant latent heat will delay the phase transition, again resulting in a
larger relic abundance for glueballs.

For these reasons, we believe that assuming that the phase transition is second order,
and neglecting any possible cannibalism in the dark sector is conservative.
We will simply estimate  the relic abundance in this regime by assuming that
the dark sector is matter dominated below the phase transition.
As discussed above, even if we assign an order of magnitude uncertainty on the
relic abundance of the dark glueballs after reheating, this does not substantially
affect our results.

\subsection{Freeze-in with Low-Energy Gluons: $\Trh \ll T_c$}
\label{subsec:abundance_T_low}

Finally, we consider the regime $\Trh \ll \Tc$.
This regime is not important for our constraints, so our
discussion will be brief.
In this regime, the reheating takes place through the 
inverse decay process
$\gamma\gamma \to G$, where $G$ is a $0^{++}$ glueball.
A quantitative estimate of the rate requires matching the 
operator in \Eq{L_IR} onto an interaction between photons 
and dark glueballs.
Roughly, we expect
\[
{\cal L}_\text{int} \sim \frac{\alpha_2 \alphad \Lambda^3}{m_\Q^4} 
G F^{\mu\nu} F_{\mu\nu} \, .
\]
The important point is that the inverse decay rate is 
exponentially suppressed by a Boltzmann factor.
Using detailed balance, we can estimate it as
\[
\dot{n}_{0^{++}} + 3H{n}_{0^{++}}  \simeq
n_{0^{++}}^\text{eq}
\Gamma(G \to \gamma\gamma)
\sim \frac{\alpha_2^2 \alphad^2 \Lambda^9}{m_\Q^8}
\left( \frac{m_{0^{++}} \Trh}{2\pi} \right)^{3/2}
e^{-m_{0^{++}}/\Trh}.
\]

\section{Cosmological Bounds on Quirks}

In the previous sections, we demonstrated that in models predicting quirk signals at colliders, the associated glueballs are (i) long-lived enough to inject energy into the SM bath after BBN (see \cref{sec:basics}), and (ii) copiously produced in the early universe via irreducible freeze-in processes, mediated by the same interactions responsible for the quirk signals (see \cref{sec:abundance}). 
In this section, we examine the cosmological constraints on glueballs and their implications for the model’s parameter space. 
These constraints impose an upper bound on the reheat temperature of the universe.
Consequently, \textit{the observation of quirky signals at colliders could provide evidence that the post-inflationary universe was reheated only to relatively low temperatures}.
We further argue that our quantitative results are highly insensitive to uncertainties in the confining dynamics.
Moreover, we argue that although additional invisible decay channels for the dark glueballs can potentially 
evade our bounds 
in the \textit{microscopic and mesoscopic} quirk regimes, the discovery of a \textit{macroscopic} quirk would still unavoidably imply a low reheat temperature -- even in the presence of such new decay channels.

\subsection{Upper Bound on the Reheat Temperature}
\label{sec:bounds}

In the benchmark models under consideration, the glueballs predominantly decay into a pair of photons.
Depending on their lifetime, $\tau_{0^{++}}$, these long-lived relics can be subject to various cosmological constraints, arising from different epochs in the early universe.
\begin{itemize}
    \item 
    For lifetimes in the range $10^2 ~\mathrm{s}\lesssim \tau_{0^{++}} \lesssim 10^{12}~\mathrm{s}$, the decay of glueballs can produce energetic photons during or after the BBN epoch, potentially disrupting the primordial abundances of light elements. For corresponding bounds on glueball abundances, see Ref.~\cite{Kawasaki:2020qxm}.

    \item For $10^5 ~\mathrm{s}\lesssim \tau_{0^{++}} \lesssim 10^{25}~\mathrm{s}$, glueball decays can leave observable imprints on the Cosmic Microwave Background (CMB). In particular, such decays can lead to detectable spectral distortions, as well as modifications to the CMB anisotropies \cite{Dimastrogiovanni:2015wvk,Slatyer:2012yq,Slatyer:2016qyl}.

    \item For $10^{17} ~\mathrm{s}\lesssim \tau_{0^{++}} \lesssim 10^{27}~\mathrm{s} $, various $\gamma$-ray searches can be sensitive to the glueball decay \cite{Essig:2013goa}.%
    \footnote{For lifetimes comparable to those probed by various $\gamma$-ray searches, glueball decays can also produce observable signals in 21 cm surveys \cite{Clark:2018ghm,Liu:2018uzy,Mitridate:2018iag}. We find that the resulting constraints are subdominant compared to other bounds relevant to our model, so we do not include them in our analysis.}  

    \item For $\tau_{0^{++}} \gtrsim 10^{17} ~\mathrm{s}$, the glueballs can act as a component of dark matter.
    This can  overclose the universe.
    
    \item 
    Cosmologically stable light glueballs have large self-interaction cross sections, which give constraints even if the glueballs are only a subcomponent of dark matter.%
    \footnote{Light and long-lived glueballs also behave like warm/hot dark matter relics; in our model, the bounds on the abundance of such relics is sub-dominant to the constraints arising from self-interaction.}
    For these bounds, we use the self-interaction cross section \cite{Forestell:2016qhc}
\begin{equation}
     \sigma_{22} v \sim \frac{(4\pi)^3}{N^4} \frac{1}{m_{0^{++}}^2},
     \label{eq:2to2xsec}
\end{equation}
and compare it to the upper limit on dark matter self-interactions, $\sigma/m \lesssim 1 \,\mathrm{cm}^2/\mathrm{g}$ - see, e.g., Ref.~\cite{Randall:2008ppe,Harvey:2015hha,Tulin:2017ara}. 
The estimate for the cross section above is only an order-of-magnitude estimate, but as discussed previously, the bounds depend on a high power of $\Trh$, so this does not stronly affect the resulting bounds.

    \item If the relic glueballs are sufficiently light, 
    the dark gluons can contribute to $N_\mathrm{eff}$ measured at CMB \cite{Planck:2018vyg,Yeh:2022heq,ParticleDataGroup:2024cfk} or BBN \cite{Yeh:2022heq}. 
    Using the conservation of entropy in dark and the visible sector separately, the dark gluon contribution to $\dNeff$ 
    is given by
\begin{equation}
    (\dNeff)_f = \frac{8}{7} \left( \frac{11}{4} \right)^{4/3} \left( \frac{g_{\text{dark},i}}{g_{\text{dark},f}} \, \frac{g_{*,f}}{g_{*,i}} \right)^{4/3} \frac{g_{*,f}}{2} \,\xi_i^4 \, ,
    \label{eq:Neff_def}
\end{equation}
where $g_{*}$ ($g_\text{dark}$) denotes the SM (dark sector) number of relativistic degrees of freedom, $\xi$ denotes the ratio of the temperatures of the two baths, and subscript $i$ ($f$) denote an arbitrary initial (final) moment in cosmological history.\footnote{See Ref.~\cite{Buen-Abad:2015ova} for a previous study of confining dark sectors contributing to $\dNeff$.}

\end{itemize}

\begin{figure}
    \centering
    \resizebox{\columnwidth}{!}{
    \includegraphics[width=0.5\linewidth]{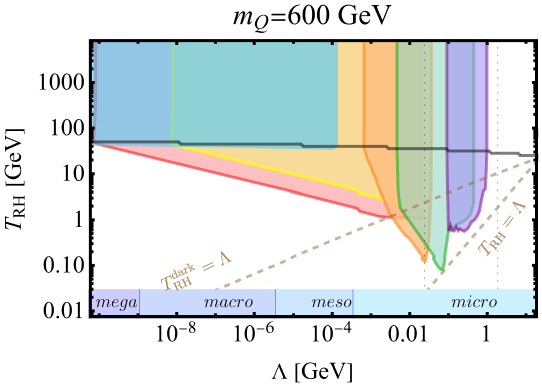}
    \hspace{0.2in}
    \includegraphics[width=0.5\linewidth]{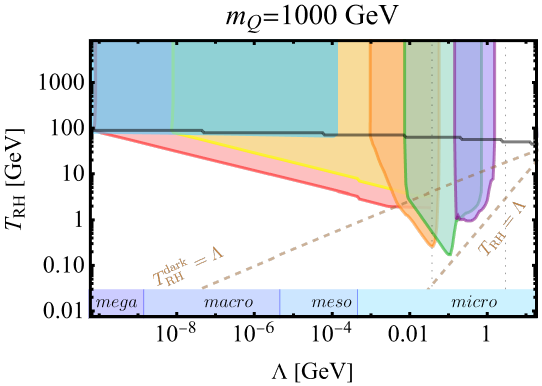}
    }\\
    \resizebox{\columnwidth}{!}{
    \includegraphics[width=0.5\linewidth]{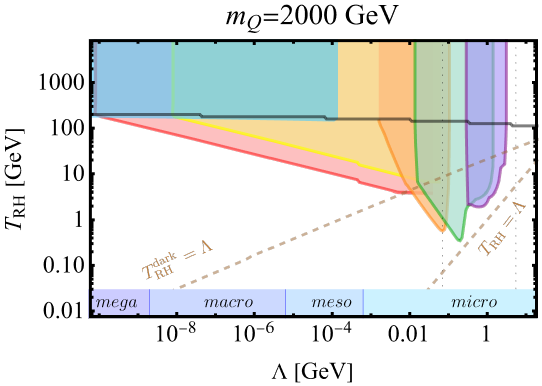}
    \hspace{0.2in}
    \includegraphics[width=0.5\linewidth]{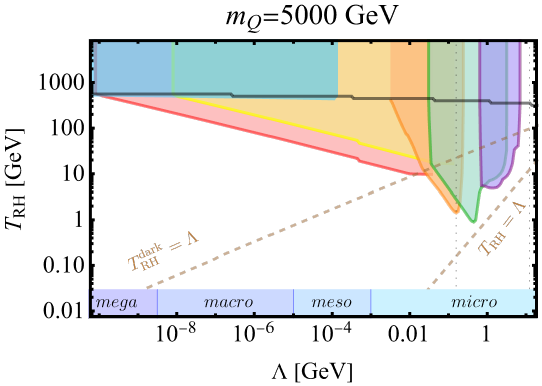}
    }\\
    \resizebox{\columnwidth}{!}{
    \includegraphics[width=0.5\linewidth]{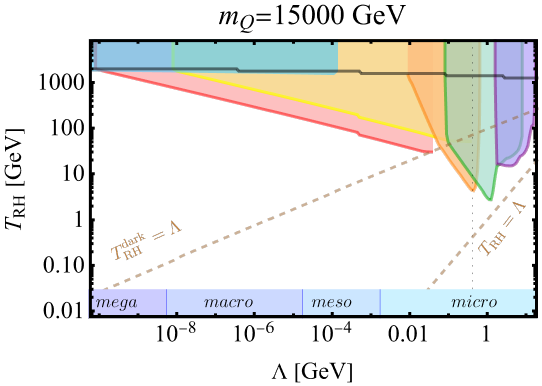}
    \hspace{0.2in}
    \includegraphics[width=0.5\linewidth]{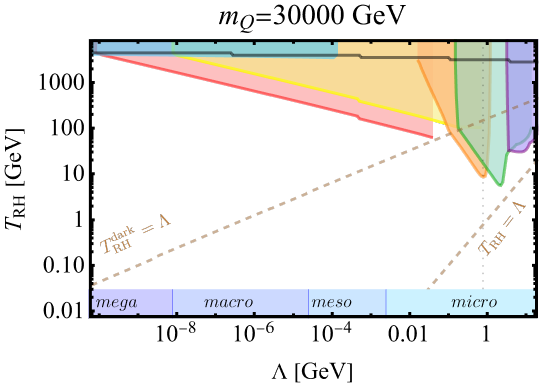}
    }
    \caption{The upper bound on the reheat temperature ($\trh$) of the universe as a function of the dark confinement scale $\Lambda$, for different values of the quirk masses $m_\Q$ (different panels). 
    Different colored regions are ruled out by constraints on long-lived glueballs from BBN \cite{Kawasaki:2020qxm} (\textcolor{custompurple}{purple}), CMB \cite{Dimastrogiovanni:2015wvk,Slatyer:2016qyl} (\textcolor{customgreen}{green}), $\gamma$-ray searches \cite{Essig:2013goa} (\textcolor{customorange}{orange}), overclosure \cite{Planck:2018vyg,ParticleDataGroup:2024cfk} (\textcolor{customyellow}{yellow}), DM self-interactions \cite{Randall:2008ppe,Harvey:2015hha,Tulin:2017ara} (\textcolor{customred}{red}), and $\dNeff$ at BBN \cite{Yeh:2022heq} (\textcolor{turquoise}{turquoise}). Above the \textcolor{black}{black} line, the two sectors reach equilibrium and the abundance of dark glueballs becomes independent of the reheat temperature. 
    Discovering a quirk at colliders for any value of $m_\Q$ and $\Lambda$ where the colored region bounds extend below the \textcolor{black}{black} line, puts an upper bound on $\trh$. 
    Different blue bars at the bottom demarcate different quirk size regions. 
    The dashed \textcolor{custombrown}{brown} lines mark milestone reheat temperatures of the dark sector, see the text for further details. The left (right) vertical dotted line marks where the glueball lifetime is equal to the age of the universe (is equal to one second). For lower values of $\Lambda$, $\dNeff$ bounds constrain the reheat temperature to around the \textcolor{black}{black} line, see the text for details. Quirks with larger values of $\Lambda$ are not constrained by cosmological observables considered here. }
    \label{fig:money}
\end{figure}

We recast the constraints from all the aforementioned analyses onto the parameter space of our model. 
The excluded regions are shown as colored regions in Fig.~\ref{fig:money} for six different benchmark quirk masses.
The glueball decays are discussed in \cref{sec:basics}.
The calculation of the relic abundance of the glueballs is described in \cref{sec:abundance} and Appendix~\ref{app:freezein}.

In each panel of Fig.~\ref{fig:money}, the \textcolor{black}{black} and \textcolor{custombrown}{dashed brown} contours partition the parameter space into four regions, corresponding to qualitatively different freeze-in histories for the glueballs:
\begin{itemize}
\item 
In the region above the solid black line, dark sector and the SM come to equilibrium below the reheat temperature, so the relic glueball abundance is independent of $\Trh$.
 This threshold is insensitive to the confinement scale $\Lambda$; the only dependence on $\Lambda$ comes from the logarithmic running of $\alphad$.
\item 
The region between the \textcolor{black}{solid black} line and the \textcolor{custombrown}{dashed brown} line labeled $\Td^\text{RH} = \Lambda$ corresponds to the high-energy gluon regime discussed in \cref{subsec:abundance_T_high}.
In this regime, the dark sector is first populated as a bath of \emph{relativistic} gluons, and then cools through the confining phase transition.

\item
The region between the \textcolor{custombrown}{dashed brown} lines labeled $\Trh^\text{dark} = \Lambda$ and $\Trh = \Lambda$ corresponds to the intermediate-energy gluon regime discussed in \cref{subsec:abundance_T_mid}.

\item 
The region below the \textcolor{custombrown}{dashed brown} line labeled $\Trh = \Lambda$ corresponds to the low-energy regime discussed in \cref{subsec:abundance_T_low}.
The figures show that this region is not relevant for our bounds.

\end{itemize}

The excluded regions are as follows:

\begin{itemize}

\item
The \textcolor{customyellow}{yellow} region shows the parameter space where the relic glueballs overclose the universe. 
On the \textcolor{customyellow}{yellow} line at the lower end of this region, the dark glueballs account for the entire dark matter abundance.

\item
The \textcolor{customred}{red} region shows the parameter space where the theory is ruled out by self-interaction constraints on dark matter.
These bounds are stronger than the overclosure bounds because they constrain the theory even if the glueballs are a subcomponent of dark matter.

\item
The \textcolor{turquoise}{turquoise} region in the upper left corner of the plots shows the parameter space where the theory is ruled out by $\dNeff$ constraints during BBN.\footnote{We use the $N_\text{eff} \simeq 2.895^{+0.142}_{-0.141}$ from BBN measurements \cite{Yeh:2022heq}, which marginally rules out the case with $N=2$ dark colors; all larger dark gauge groups are definitely ruled out as the contribution of dark gluons to $\dNeff$ scales as $N^2$.  }
This constraint persists even for confinement scales $\Lambda$ smaller than those displayed in \cref{fig:money}.
However, in this regime quirks are indistinguishable from ordinary heavy stable particles at colliders.

\item The remaining \textcolor{customorange}{orange}, \textcolor{customgreen}{green}, and \textcolor{custompurple}{purple} regions denote constraints arising from glueball decays, as probed by $\gamma$-ray searches, the CMB, and BBN, respectively.
    
\end{itemize}

It should be noted that our abundance calculation neglects the
effect of the inverse process $g'g' \to \gamma\gamma$.
This introduces some error as the two sectors approach equilibrium, {\it i.e.}~near the black line in \cref{fig:money}. 
This, in turn, introduces a small error in the exact $\Lambda$ value where different colored regions intersect the equilibrium line in \cref{fig:money}.

The main takeaway from these results is that in the minimal quirk model under study, \textit{discovering quirks at colliders, and in a large range of $m_\Q$ and $\Lambda$ values, implies the reheat temperature of the universe $\trh$ is bounded from above.} 
In the plots, this corresponds to the parts of the parameter space where the cosmological constraints (colored regions) extend below the \textcolor{black}{black} line. 
The existence of this upper bound, lying in the range $ \Trh \lesssim \mathcal{O}(10^{-1}-10^3)$~GeV for collider-detectable quirks, constitutes the main result of this paper.

We reiterate that in the \textit{mesoscopic} and \textit{macroscopic} quirk regimes, the signals predicted by this model are highly distinctive and cannot be mimicked by any other known BSM scenarios. 
Thus, the observation of such signals, when combined with our results in \cref{fig:money}, would strongly point to the existence of an upper bound on the reheating temperature.
By contrast, despite their phenomenological richness, signals arising from \textit{microscopic} quirks can be reproduced by other BSM models. 
Therefore, in the event of discovering such a signal, one should exercise caution in inferring an upper bound on the reheating temperature of the Universe.

Remarkably, we see that the cosmological bounds essentially cover the entire region of the parameter space of the minimal quirk model that is of interest to present and planned colliders.
Next,
we will comment on the robustness and model-dependence of these bounds.

\subsection{Comments on Uncertainties}
\label{subsec:uncertain}

In this section we comment on the sensitivity of our bounds to various model assumptions and theoretical uncertainties.
Specifically, we comment on our assumptions about the reheating mechanism, and uncertainties arising from the strongly coupled
nature of the dark sector.
Some of these comments have been made earlier, but we collect them here for convenience.

We begin by discussing our assumptions regarding reheating.
We define the reheat temperature $\Trh$ to be the highest temperature of the SM in the most recent radiation-dominated era.
This is not necessarily the same as the reheat temperature after inflation;
for example there may be a late matter-dominated era (e.g., due to the out-of-equilibrium decay of a heavy particle)
after inflationary reheating that ends with another reheating era.
Our bounds are obtained by assuming that the dark sector is not directly populated by the most recent reheating mechanism.
Because the quirks are charged under the SM gauge group, we must also assume that $\Trh \lesssim m_\Q/25$ to avoid equilibrating the quirks, and hence the dark gluons.

A reheating sector satisfying these assumptions can be realized physically 
by a sector that reheats the standard model instantaneously ({\it i.e.}~on a time scale rapid
compared to $H$), and which has negligible couplings to the quirks and dark gluons.
However, there are many other possibilities for reheating. 
The simplest is that the reheating is due to the decay of a heavy stable particle or the coherent oscillations of a field that decays slowly.
In such a scenario, the SM reaches temperatures \emph{above} $\Trh$ during the reheating process, and the UV freeze-in mechanism discussed in this paper will give a \emph{larger} relic abundance to the dark sector.
More generally, it is clear that the dynamics above the reheat temperature can only increase the relic abundance of the dark sector compared to our assumptions.
This may be interesting because the constraints in this paper require that the reheating temperature is low,
which in turn calls for non-conventional mechanisms to explain baryogenesis, e.g. see models in Refs.~\cite{Ghalsasi:2015mxa,Aitken:2017wie,Elor:2018twp,Nelson:2019fln,Elahi:2021jia}.
If these rely on non-minimal reheating mechanisms, then one must check that these do not populate the dark sector in this model, resulting in stronger cosmological constraints from the dark gluons.

Next, we comment on the uncertainties due to the dark phase transition.
In the region where $T^{\rm RH}_{\rm dark} > \Lambda$, the dark sector reheats above the confinement temperature, and then cools through the phase transition.
In this case, the uncertainty comes from the cosmological evolution of the dark sector through the phase transition.
Assuming that the phase transition is second order (known to be the case for SU(2) gauge theory), the evolution through the phase transition is expected to be adiabatic, so we need to know the equation of state of pure Yang-Mills theory near the phase transition.
Lattice data indicates that this deviates significantly from that of a free gluon gas well above $\Tc$ \cite{Boyd:1996bx}.
We are not aware of any lattice data giving precise information about the equation of state below $\Tc$, but it is possible that this regime includes a period of `cannibalism' where $2 \leftrightarrow 3$ glueball interactions result in a slow cooling of the dark sector.
We have simply assigned a generous theoretical uncertainty to this evolution, using the bracketing procedure described in \cref{sec:abundance}.

For SU($N$) gauge group with $N > 2$, the phase transition is first order.
In this case, there are additional complications and theoretical uncertainties.
One important effect is that in the adiabatic limit the expansion of the universe must remove the latent heat from the dark sector in order to complete the transition.
This results in a period of expansion where the temperature of the dark sector does not change,
which tends to \emph{increase} the relic abundance.

In addition, the bubble nucleation rate may be suppressed in a first order phase transition, leading to a period of supercooling.
In this case, the dark sector redshifts like radiation even after the phase transition, which tends to \emph{decrease} the final relic abundance.
Recent lattice studies suggest that supercooling is a small effect for moderate values of $N$.
In the thin wall approximation, the Euclidean action for bubble nucleation near $T = \Tc$ is given by (see the appendix in Ref.~\cite{Asadi:2021pwo} and references therein)
\[
S = \frac{16\pi}{3} \frac{\sigma_{\mathrm{surf}}^3}{\ell_\text{heat}^2} 
\frac{T_c^2}{(T-\Tc)^2} \, ,
\label{eq:bubble_S}
\]
where $\sigma_{\mathrm{surf}}$ is the surface tension and $\ell_\text{heat}$ is the latent heat.
The lattice measurements of \cite{Rindlisbacher:2025dqw} find
\[
\frac{\sigma_{\mathrm{surf}}}{\Tc^3} \simeq  0.018 N^2 - 0.19,
\qquad
\frac{\ell_\text{heat}}{\Tc^4} \simeq 0.36 N^2 -1.9.
\]
Because the bubble nucleation is exponentially sensitive to the action in \Eq{bubble_S}, this
suggests an extended supercooling period exists only for $N \gtrsim 40$.
Such large values of $N$ will give rise to a Landau pole for the SM gauge couplings near the quirk mass.
In addition, for large $N$ and sufficiently small $m_\Q$ there will be additional constraints from precision electroweak tests, e.g.~the $W$ parameter
\cite{Barbieri:2004qk,Cacciapaglia:2006pk,Alves:2014cda,Farina:2016rws}.
However, we are estimating an exponent, and the use of the thin-wall approximation is not obviously justified.
What is robust is that the lattice results tell us that for moderate values of $N$ the supercooling is not a large effect, and in this case it seems reasonable to assign an uncertainty of an order of magnitude in the glueball relic abundance. 

The uncertainties discussed above result in only a minor change in the inferred bound on $\Trh$, because the glueball relic abundance scales steeply with the reheat temperature ($\Omega h^2 \sim \Trh^7$).
We find that a factor of 100 uncertainty in the glueball abundance translates into at most a factor of 2 in the bounds on $\Trh$.

Similar comments apply to the uncertainty in the decay rate of the lightest glueball.
We used a simple estimate for the decay rate, see Eq.~\eqref{eq:decay_width}.
An order of magnitude uncertainty on this decay rate moves the bounds on $\Trh$ to larger or smaller values of $\Lambda$ merely by an order-1 factor. 

Finally, we comment briefly on the possible effects of heavier glueballs.
As already discussed above, if $\Td^\text{RH} > \Tc$, the $3 \leftrightarrow 2$ interactions among glueballs will strongly suppress the relic abundance of the heavier glueballs.
However, in the regime where the dark sector is directly reheated to a temperature below $\Tc$, the $3 \leftrightarrow 2$ interactions are never in equilibrium, and it is possible that the abundance of the heavier dark gluons is not very suppressed compared to that of the $0^{++}$ states.
(For example, it seems reasonable that the abundance of glueballs with mass $m$ is proportional to the Boltzman factor $e^{-m/\Td^\text{RH}}$.)
We have not investigated the decays of the heavier states, but they may give additional constraints if some of them have lifetimes significantly longer than that of the $0^{++}$ states.
However, if this is the case, then these will give additional constraints, so ignoring these effects is a conservative assumption.

\subsection{Beyond the Minimal Quirk Model}

In this section we consider various extensions of the benchmark quirk model and comment on the implications for the cosmological bounds.

\subsubsection{QCD Colored Quirks}
Another natural extension of the minimal model is that the quirks are charged under QCD color as well as electromagnetism.
In this case, integrating out the quirks will generate an additional dimension-8 interaction
\[
\Delta{\cal L} \sim  \frac{g_s^2 g'^2}{16\pi^2 m_\Q^4} \tr(G'^{\mu\nu} G'_{\mu\nu}) 
\tr(G^{\rho\sigma} G_{\rho\sigma}),
\]
where $G_{\mu\nu}$ is the QCD color field strength and $g_s$ the QCD gauge coupling.
This will enhance the relic abundance of dark glueballs through the process $gg \to g'g'$, where $g$ is a SM gluon.
This change, by itself, will not strongly affect the bounds
because they are very insenstive to the precise relic abundance, as discussed above.
The operator above will also allow the dark glueballs to decay to SM hadrons if the dark glueball mass is larger than $2m_\pi$.
In this case, the decays to hadrons reduce the lifetime of the dark glueballs, and shift the bounds in \cref{fig:money} to slightly lower values of $\Lambda$.
However, because the lifetime depends so steeply on $\Lambda$ (see \cref{fig:lifetime} and \Eq{decay_width}), this shift is expected to be rather small. 
One significant change in this case is that the BBN bounds may extend to lifetimes as short as $\sim 1$~sec \cite{Jedamzik:2006xz}.

Next, we consider the case where the quirks are not electrically charged, but carry SM color.
In this case, the gluon lifetime becomes significantly longer if $m_{0^{++}} < 2m_\pi$, because decays to hadrons are kinematically forbidden, and the dark glueballs can decay only to leptons and photons through higher-loop processes.
For these values of $\Lambda$ ($\Lambda = T_c \gtrsim 50$~MeV in our baseline model), the bounds in \cref{fig:money} are dominated by constraints on cosmologically stable glueballs, and are again unmodified.
The precise bounds for larger values of $\Lambda$ will change somewhat due to the the different lifetimes and decay modes as discussed above, but we will leave investigation of this for future work.

\subsubsection{Quirks with Yukawa Couplings}
Another possible extension of the minimal model is to have Yukawa couplings of the quirks
to the SM Higgs field.
For example, the quirks could have the SM quantum numbers of a vector-like generation of leptons.
This allows an electroweak preserving mass for the quirks in addition to Yukawa couplings to the Higgs.
If we assume that the electroweak preserving mass dominates, this is much less constrained by precision electroweak tests.
In this case, integrating out the quirks will induce dimension-6 operators of the form
\[
\Delta {\cal L} \sim \frac{y_Q^2}{16\pi^2 m_\Q^2}
H^\dagger H \tr(G'^{\mu\nu} G'_{\mu\nu}),
\]
where $y_Q$ is the quirk Yukawa coupling.

This can modify the cosmological bounds on dark glueballs in several ways.
First, there is a 2-loop contribution to the dimension-8 operator in the effective field theory below the Higgs mass of order
\[
\Delta{\cal L} \sim \frac{e^2 y_Q^2}{(4\pi)^4}
\frac{1}{m_\Q^2 m_h^2} 
\tr(G'^{\mu\nu} G'_{\mu\nu})
F^{\rho\sigma} F_{\rho\sigma},
\]
which is larger than the 1-loop contribution \Eq{L_IR} if $m_\Q \gtrsim 4\pi m_h / y_Q$.
If this is the case, the bounds for this model will be stronger than the bounds for our minimal model with the same values of $m_\Q$ and $\Lambda$.

The effective theory below the Higgs mass will also contain operators coupling the dark gluons to SM fermions, such as
\[
\Delta{\cal L} \sim \frac{y_Q^2 m_e}{(4\pi)^2 m_\Q^2 m_h^2}
\tr(G'^{\mu\nu} G'_{\mu\nu})
\bar{e} e.
\]
As a result, the process $e\bar{e} \to g'g'$ will dominate the freeze-in for $m_\Q \gtrsim m_h (\Trh/m_e)^{1/2}(e/y_Q)$.
If this is the case, it will increase the relic abundance, which again strengthens the bounds.

In addition, the decay $0^{++} \to e\bar{e}$ may be important if $m_{0^{++}} > 2m_e$.
This occurs only for the largest values of $\Lambda$ that are relevant for colliders.
If this is significant, the main effect is to shift the excluded regions toward smaller values of $\Lambda$, but the bounds will be qualitatively similar.

\subsubsection{Non-Minimal Models with Invisible Glueball Decays}

One can ask what other types of modifications to the model we have considered would be sufficient to (potentially) avoid the strong bounds on $T_{\rm RH}$ that we have found.  
Figure~\ref{fig:money} clearly demonstrates that specific constraints  arise from three distinct underlying sources:
\begin{enumerate}
\item Glueball decays to photons give strong constraints for glueball lifetimes between the timescale of about BBN to nearly $10^{10}$ times the age of the universe.
\item Glueballs that live longer than the age of the universe can contribute to a sub-component of dark matter that have strong constraints from their large self-interactions.
\item Dark gluons contribute to $\Delta N_{\rm eff}$ that is constrained by BBN measurements so long as the dark sector is still in the deconfined phase.
\end{enumerate}
This suggests that if we modify the minimal quirk model theory with i) additional very light particles that are completely neutral under both the SM and the dark gauge group, and ii) new interactions of dark gluons with the light particles that \emph{significantly exceed} the interactions of dark gluons with the SM, then it could be possible to evade 
the decay constraints 
by having glueballs decay invisibly.
A more rapid decay of glueballs will also change the regime where glueballs live long enough to contribute to a self-interacting contribution of dark matter.

Nevertheless, even in the presence of invisible decays, constraints remain that are straightforward to see independently of the particular model implementation:
\begin{itemize}
\item If the dark sector reaches thermal equilibrium with the SM (above the black line in Fig.~\ref{fig:money}), we have seen there is a constraint from
$\Delta N_{\rm eff}$ that arises if the dark gluons have not yet confined by the time of BBN, i.e., $\Lambda \lesssim 10^{-4}$~GeV\@. This constraint obviously remains even if after confinement the dark glueballs decay invisibly, and thus the \textcolor{turquoise}{turquoise} region above the black line in Fig.~\ref{fig:money} remains a robust constraint on $T_{\rm RH}$.
\item The freeze-in constraints
on glueballs abundance when they have a lifetime longer than the age of the universe also persist.  The actual region of parameter space that is constrained depends on recalculating the glueball lifetime in the presence of the light states, but lifetimes cannot be made arbitrarily short since one needs at least a dimension-6 (dimension-7) operator involving the dark gluons if the light states are scalars (fermions).
\item New constraints can arise at larger confinement scales, in the regime where the dark sector reaches thermal equilibrium.  In this case, the dark glueballs decay to the light sector states before BBN, and these light states contribute excessively to $\Delta N_{\rm eff}$.
\end{itemize}
In addition, if even a very small abundance of excited glueballs states remains,
and if these excited glueballs can only decay through dimension-8 operators (e.g., $0^{-+}$ \cite{Juknevich:2009gg}),
additional constraints may arise even if the lightest glueball predominantly decays to invisible states.

In Appendix~\ref{app:appendage}, we provide an example of these generic features that arises when we add a massless scalar field coupled to dark gluons through a dimension-6 operator. 
This example model serves as a proof by construction that the upper bound on $\Trh$ can be evaded in case of discovering \textit{microscopic} or \textit{mesoscopic} quirks, though at the cost of considerable non-minimality of the model with new interactions and new light neutral fields entirely disconnnected from the SM and the dark sector.
However, as the example provided in the appendix shows, discovering \textit{macroscopic} (and megascopic) quirks still continues to enforce stringent constraints on $\Trh$. 
This arises from bounds on $\Delta N_\text{eff}$ due to dark gluons, precisely as outlined above.

\section{Glueball Dark Matter}
\label{subsec:GB_DM}

An interesting observation from \cref{fig:money} is that there is a small window of $\Lambda$ and $m_\Q$ masses for which the glueballs in these quirk models can explain the entirety of dark matter today. 
This is the region where the yellow line ($\Omega_{0^{++}}h^2 \sim 0.11$) is not ruled out by the \textcolor{customred}{red} (self-interaction) or the \textcolor{customorange}{orange} ($\gamma$-ray) bounds. 
In this section, we will study the possibility of dark glueballs constituting a non-negligible fraction of DM in quirk 
models. 

Dark glueballs can only be viable dark matter candidates if the coupling between the dark sector and the SM is sufficiently small to render them long-lived on cosmological timescales. This requirement effectively pushes glueball dark matter toward the quirk regime where the interaction between the glueballs and the SM is suppressed due to the heavy mass of the particles charged under the dark gauge group.
As shown in the previous section, in this limit, the dark gluons are populated via UV freeze-in. 
In this section, we use the results of the previous sections to delineate the viable parameter space for such glueball dark matter scenarios.

We continue to focus on the benchmark models introduced in \cref{sec:basics}, while noting that alternative portal choices do not significantly alter the glueball lifetime or the relic abundance calculation. The free parameters of the model are the dark confinement scale $\Lambda$, the quirk mass $m_\Q$, and the reheat temperature $\trh$. 
By requiring that the relic abundance of dark glueball dark matter matches the observed value today \cite{Planck:2018vyg,ParticleDataGroup:2024cfk}, we fix $\trh$ in terms of the other two parameters.

\begin{figure}
    \centering
    \includegraphics[width=0.65\linewidth]{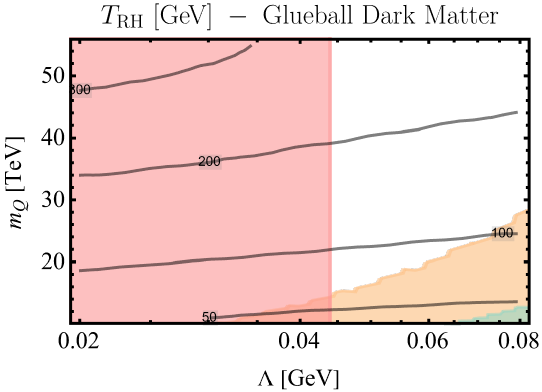}
    \caption{The viable range of dark confinement scale $\Lambda$ and  quirk masses $m_\Q$ in models of non-secluded dark glueball dark matter. We show the required $\trh$ (black) such that the glueballs have the right DM relic abundance today. Only the white part of the parameter space is viable thanks to the bounds on decaying DM lifetime from CMB (\textcolor{customgreen}{green}) or from $\gamma$-ray (\textcolor{customorange}{orange}), as well as self-interaction constraints (\textcolor{customred}{red}); future updates on these quantities can probe this parameter space. 
    }
    \label{fig:GB_DM}
\end{figure}

In Fig.~\ref{fig:GB_DM} we scan over different values of $\Lambda$ and $m_\Q$ for this benchmark setup. 
For each value, we calculate the reheat temperature that gives rise to the correct DM abundance today. 
Higher $\trh$ values overclose the universe for each point on this plot. 
Note that requiring long enough lifetimes to evade $\gamma$-ray bounds puts a lower bound on $m_\Q/\Lambda$, while the self-interaction bounds put a lower bound on the value of $\Lambda$. 
The combination of these bounds pushes the viable part of the parameter space for non-secluded glueball DM to $\Lambda \gtrsim \mathcal{O}(40)$~MeV and quirk masses $m_\Q \gtrsim \mathcal{O}(10)$~TeV, i.e. beyond the reach of near-future colliders. 

As a result, we can argue that quirk signals (with $m_\Q \lesssim \mathcal{O}(10)$~TeV) and non-secluded glueball DM are at odds with each other. 
Alternatively, this plot implies if non-secluded glueballs are found to constitute the majority of dark matter today, the associated quirk masses must lie in the range $m_\Q \gtrsim \mathcal{O}(10)$~TeV.
Changing some details of the model, e.g. number of colors or the coupling between the quirks and SM, could potentially change the threshold $m_\Q$ above, but will leave this qualitative conclusion intact. 
This figure also underscores the fact that future improvements on DM lifetime and its self-interaction can probe parts of the viable parameter space of non-secluded dark glueball DM models.
In this treatment we have neglected various uncertainties pertaining to the dark confinement dynamics, see the discussion in \cref{subsec:uncertain};
the exact location of regions and contours in \cref{fig:GB_DM} is subject to these uncertainties, but our qualitative conclusions remain.

\section{Discussion and Conclusions}

In this paper, we have studied the cosmological implications of quirks, new heavy fermions coupled to a new confining SU($N$) gauge group
that also transform under the SM gauge group.  
The minimal quirk model is fully specified by the quirk mass $m_\Q$, the confinement scale $\Lambda$, and the SM quantum numbers of the quirks.
This simple model can give rise to a plethora of exotic signals at colliders because the quirks are connected by a color flux string that does not break.
We find that in the regime where the flux string can give observable effects at current or planned colliders, cosmological constraints imply that the reheat temperature is below the electroweak scale:
\begin{center}
  \begin{tcolorbox}[colframe=custompurple, colback=white, width=\linewidth, boxrule=2pt,halign=center]
    \textbf{In minimal models,
    the observation of `quirky' signals at colliders implies} \mbox{$\trh~\lesssim~\mathcal{O}(100)$}~\textbf{GeV}.
  \end{tcolorbox}
\end{center}
The bound arises because the glueballs of the new strong interactions are long-lived, and have an irreducible relic abundance from freeze-in.
The glueballs are very light; their decays to photons are constrained by  BBN, the CMB, and astrophysical observations of $\gamma$-rays.
If the lifetime is longer than the age of the universe, additional constraints arise from the glueballs making up too large a component of self-interacting dark matter.

The freeze-in glueball abundance has a strong dependence on $\Trh$ arising from the high dimension (dimension-8) of the operator that connects the SM to the dark sector.  This implies that many of the uncertainties intrinsic to the strongly-coupled cross sections, decay rates, and the confinement transition lead to only rather small changes to the numerical bounds and thus do not affect our conclusions.  
Moreover, if reheating is non-instantaneous, the maximum SM temperature can exceed $\trh$, enhancing the glueball abundance and strengthening the upper bound on the reheat temperature. 

We have also showed that our results are robust against variations of the quirk model, including the possibility that quirks transform under the QCD, as well as quirks having Yukawa couplings to the Higgs boson.  These model variations can significantly change the collider physics of quirks, but the bounds on $\Trh$ remain largely unchanged.

We also investigated the possibility of attempting to avoid the constraints by introducing new neutral light states that can dominate the decays of the glueballs.
This also requires the introduction of additional interactions between the new light states and the quirk sector that require UV completion not tremendously far from the TeV scale.
Even with this modification, we find that the $\Delta N_{\rm eff}$ bounds from BBN remain for the regime where the quirk strings are macroscopic, because these bounds are insensitive to how the dark glueballs decay:
\begin{center}
  \begin{tcolorbox}[colframe=customorange, colback=white, width=\linewidth, boxrule=2pt,halign=center]
    \textbf{In non-minimal models where glueballs decay to additional light states,
    the observation of \emph{macroscopic} `quirky' signals
    implies $\trh \lesssim \mathcal{O}(100)$~\textbf{GeV}.}
  \end{tcolorbox}
\end{center}

We found that for a large portion of the quirk parameter space shown in Fig.~\ref{fig:money}, the upper bound on the reheat temperature is $\trh \lesssim 100$~GeV, i.e. the electroweak symmetry breaking scale. 
For this part of the parameter space, most baryogenesis models, including leptogenesis and electroweak baryogenesis, are excluded, highlighting the need for novel models to explain the observed asymmetry of SM baryons (as well as the coincidence problem).

A corollary of our results is that glueball dark matter is viable only in a rather esoteric region of the quirky parameter space.  
If glueballs were found to constitute the majority of dark matter today, the associated quirk masses must lie in the range $m_\Q \gtrsim \mathcal{O}(10)$~TeV with $\Lambda \gtrsim 0.05$~GeV, where the latter is set by the current self-interaction constraints.

Our results strongly motivate renewed collider searches for quirks, as their discovery would signal, not only new dark sectors, but also a non-standard cosmological history with a low reheat temperature.
This underscores the importance of dedicated search strategies. 
Our findings also exemplify the deep interconnections within confining dark sectors, whereby any new physics discovery can have profound
implications for the early universe cosmology.

\section*{Acknowledgements}

We thank 
David Curtin,
Joshua Forsyth, 
Stefania Gori,
Simon Knapen,
Nick Rodd,
John Terning,
Jorinde van de Vis,
and Chris Verhaaren
for illuminating discussions. 
We also thank Jessica Howard for collaboration in the early stages of this work.
The work of PA is supported in part by the US
Department of Energy grant number DE-SC0010107.
GK is supported in part by the US Department of Energy under grant number DE-SC0011640.
The work of ML is supported in part by the US Department of Energy under grant DESC0009999.
This research was also supported in part by grant NSF PHY-2309135 to the Kavli Institute for Theoretical Physics (KITP).

\appendix

\section{Dark Gluon Freeze-in}
\label{app:freezein}

In this appendix, we present the detailed calculation of the dark gluon
abundance due to freeze-in.
Although the final glueball relic abundance is subject to an order-1 
uncertainty due to the uncertainty in the treatment of the confinement 
phase transition, we will see that there are large numerical factors in
this calculation that are important.

We assume that just after reheating, the universe starts with the SM in 
equilibrium at the temperature $\Tc \ll \Trh \ll m_\Q$, while the dark gluons
are not populated.
The dark gluons are then
populated by the interaction $\gamma\gamma \to g'g'$
mediated by a dimension-8 operator
similar to Eq.~\eqref{eq:L_IR}, but with the electroweak field strength $W_{\mu\nu}$ replaced with the electromagnetic field strength $F_{\mu\nu}$. The Wilson coefficient of the operator can be quantified as \cite{Juknevich:2009ji}
\[
\frac{1}{M^4} = \frac{e^2 g^2 \tr(Q^2)}{60(4\pi)^2 m_\Q^4},
\]
where $Q$ denotes the electric charge of the quirk $\Q$. 
We write the Boltzmann equation in terms of the gluon energy density:
\[
\dotdarkrho + 4 H \darkrho = 
\frac{1}4 \int (dp_1) \cdots (dp_4)
(2\pi)^4 \delta^4(P_i - P_f)
(E_1 + E_2)
f_1 f_2 \big| \widebar{\cal M}(\gamma_1 \gamma_2 \to g'_3 g'_4) \big|^2,
\]
where $\widebar{\cal M}$ is the scattering amplitude averaged over initial spins and summed over final spins and colors.
The factor of $\frac 14$  takes into account that the initial and final states each have 2 identical particles, and
\[
(dp) = \frac{d^4 p}{(2\pi)^3} \delta(p^2) \theta(p^0)
= \frac{d^3 p}{(2\pi)^3} \frac{1}{2|\vec{p}\,|}
\]
is the Lorentz invariant integral over the mass shell for a massless particle.
We are neglecting the inverse reaction $g'g' \to \gamma\gamma$, which is
justified as long as the final dark gluon abundance is small compared to
its equilibrium value.

We write the Boltzmann equation in terms of the dimensionless variables
\[
Z = \frac{\darkrho}{\rho_\gamma},
\qquad
x = \frac{T_\text{RH}}{T} \, ,
\]
where $T$ is the SM bath temperature and $\rho_\gamma$ is the photon energy density.
This gives
\[
\frac{dZ}{dx} = \frac{C}{H \rho_\gamma x},
\]
where the collision term is
\[
C = \frac 14 \int (dp_1) \cdots (dp_4)
(2\pi)^4 \delta^4(P_i - P_f)
(E_1 + E_2) f_1 f_2
\big| \widebar{\cal M}(\gamma_1 \gamma_2 \to g'_3 g'_4) \big|^2.
\]
We can use equilibrium values for $f_{1,2}$ and $\rho_\gamma$:
\[
\rho_\gamma = 2 \times \frac{\pi^2}{30} T^4,
\qquad
f_{1,2}(|\vec{p}\,|) = \frac{1}{e^{|\vec{p}|/T} - 1}.
\]
A standard calculation gives
\[
\big| \widebar{\cal M}(\gamma \gamma \to g' g') \big|^2
= \frac{36(N^2 - 1)}{M^8} s^4
\]
for SU(N) gluons.

The collision term can be written as
\[
C = \int (dp_1) (dp_2) (E_1 + E_2) f_1 f_2  \tilde{C}(s)\, ,
\]
where
\[
\tilde{C}(s) = \frac{9(N^2 - 1)s^4}{M^8} \int (dp_3) (dp_4)
(2\pi)^4 \delta^4(p_3 + p_4 - P_i) 
=  \frac{9(N^2 - 1)s^4}{M^8} \frac{1}{2\pi}.
\]
This gives
\[
C = \frac{9(N^2 - 1)}{2\pi M^8} 
\int \frac{d^3 p_1}{(2\pi)^3} \frac{1}{2|\vec{p}_1|} 
\frac{d^3 p_2}{(2\pi)^3} \frac{1}{2|\vec{p}_2|} 
(|\vec{p}_1| + |\vec{p}_2|) 
\big[ 2 |\vec{p}_1| |\vec{p}_2| (1 - \cos\theta) \big]^4
f(|\vec{p}_1|) f(|\vec{p}_2|) \, ,
\]
where $\theta$ is the scattering angle.
The angular integral is performed using
\[
\int_{-1}^1 dc \,(1 - c)^4 = \frac{32}{5},
\]
and the remaining integrals 
can be performed exactly to obtain
\[
C = \frac{9(N^2 - 1)}{2\pi M^8} 
\frac{2^6}{5\pi^4} 
\int_0^\infty \!\! dp_1 \int_0^\infty \!\! dp_2 
 (p_1 + p_2)  p_1^5 p_2^5  f(p_1) f(p_2)
 = \frac{288\pi \zeta(7)}{5}
\frac{(N^2 - 1)T^{13}}{M^8}.
\]

To solve the rate equation, we use
\[
H \rho_\gamma = 1.09  \frac{g_*^{1/2} \Trh^6}{M_\text{Pl}} x^{-5},
\]
which gives
\[
\frac{dZ}{dx} = 167 \frac{(N^2 - 1) M_\text{Pl} T_\text{RH}^7}{g_*^{1/2} M^8}
x^{-8}.
\]
The solution is 
\[
Z_f =  167 
\frac{(N^2 - 1) M_\text{Pl} T_\text{RH}^7}{g_*^{1/2} M^8} 
\frac{1 - x_f^{-7}}{7}. 
\]
To obtain the late time abundance, we set $x_f = 0$.

\section{Constraints in the Presence of Invisible Decays}
\label{app:appendage}

Here we consider a simple toy model in which the dark gluons couple to an additional completely neutral scalar field that we assume has 
no interactions with the SM\@.
We take the scalar field to be much lighter than the glueball throughout the parameter space that we explore here:\footnote{Absent further structure, such a scalar would suffer from a severe naturalness problem, but we simply assume its mass is fine-tuned as needed since this construction merely serves as a proof of principle of glueballs decaying to new light degrees of freedom. Similar conclusions can be reached with a very light (or massless) neutral fermion.}
\begin{equation}
    \mathcal{L} \supset \frac{1}{\widetilde{M}^2} |\phi|^2 \ G_a'^{\ \mu\nu} \ G'_{\mu\nu \ a} \, .
    \label{eq:loophole_L}
\end{equation}
This operator could arise in the presence of yet additional heavy scalar fields that transform under the dark gauge group which also have renormalizable couplings with the neutral scalar field.  Hence, $\widetilde{M}$ is an effective scale that encapsulates any loop-suppression and couplings appropriate to the UV completion. 
Some heavier glueballs cannot decay through the dimension-6 operator \cite{Juknevich:2009gg} in Eq.~\eqref{eq:loophole_L}.
Whether they change our conclusion here depends on their relic abundance and is subject to various strong dynamics uncertainties. This may require additional interactions, but we will not attempt to address that here.

We can check that for 
\begin{equation}
\widetilde{M} \lesssim m_\Q \frac{m_\Q}{\Lambda} \, , 
    \label{eq:Mrange}
\end{equation}
this decay channel to new light neutral scalar field $\phi$ will dominate over the decay rate to the SM\@.

\begin{figure}
    \centering
    \resizebox{\columnwidth}{!}{
    \includegraphics[width=0.5\linewidth]{figures/money_600_20.pdf}
    \hspace{0.2in}
    \includegraphics[width=0.5\linewidth]{figures/money_30000_20.pdf}
    }\\
    \resizebox{\columnwidth}{!}{
    \includegraphics[width=0.5\linewidth]{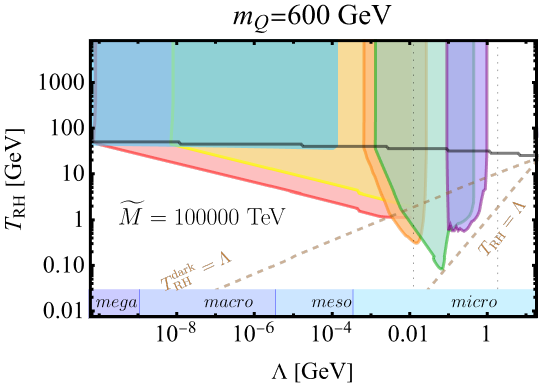}
    \hspace{0.2in}
    \includegraphics[width=0.5\linewidth]{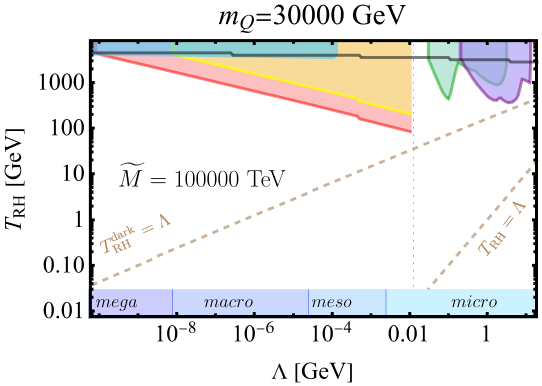}
    }\\
    \resizebox{\columnwidth}{!}{
    \includegraphics[width=0.5\linewidth]{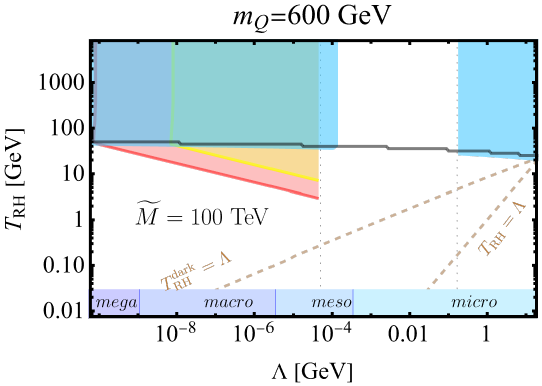}
    \hspace{0.2in}
    \includegraphics[width=0.5\linewidth]{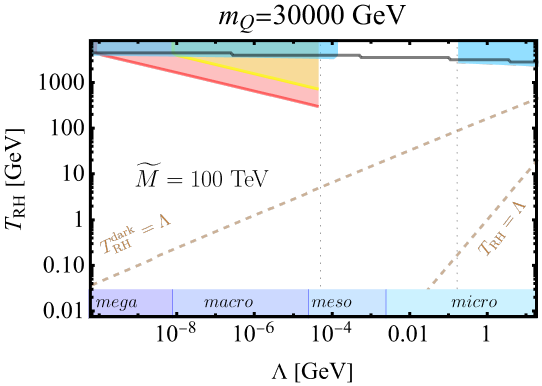}
    }
    \caption{ Same as \cref{fig:money} for two quirk masses ($m_\Q = 0.6$~TeV (\textbf{left}) and $m_\Q = 30$~TeV (\textbf{right})), but now with a new invisible decay channel for dark glueballs, see \Eq{loophole_L}. 
    Different rows show the result for different values of $\widetilde{M}$: $\widetilde{M} \rightarrow \infty$ (\textbf{top}), $\widetilde{M}=10^5$~TeV (\textbf{middle}), and $\widetilde{M}=10^2$~TeV (\textbf{bottom}). 
    The left (right) vertical dotted line marks where the glueball lifetime is equal to the age of the universe (is equal to one second). 
    As $\widetilde{M}$ decreases, the invisible decay channel dominates. 
    This results in (i) shortening glueballs lifetime and shifting the bounds to lower $\Lambda$, and (ii) loosening the BBN, CMB, and $\gamma$-ray bounds as the branching ratio to the visible channel decreases.
    The overclosure (\textcolor{customyellow}{yellow}) and DM self-interactions (\textcolor{customred}{red}) bounds are included only where glueball lifetime exceeds the age of the universe.
    Assuming the invisible product $\phi$ is massless, the $N_\mathrm{eff}$ bounds (\textcolor{turquoise}{turquoise}) now constrain higher values of $\Lambda$, where the dark glueballs decay to $\phi$ and contribute to $\dNeff$. 
    We find that such a new channel can evade the bound on the reheat temperature in case of discovering a microscopic or a mesoscopic quirk signal with specific $\Lambda$ values (where the \textcolor{black}{black} is not ruled out by colored regions), see the text for details.  }
    \label{fig:loophole}
\end{figure}

Now, we repeat the analysis of Sec.~\ref{sec:bounds} for three representative values of $\widetilde{M}$, with the results shown in \cref{fig:loophole}. The presence of the new invisible decay channel modifies the scaling of the glueball lifetime with $\Lambda$, leading to the deformation of the \textcolor{custompurple}{purple}, \textcolor{customgreen}{green}, and \textcolor{customorange}{orange} regions as $\widetilde{M}$ is varied. 
The top row corresponds to the previous case with no new invisible decay channel. 
Comparing this to other rows, we observe that the invisible channel significantly weakens the existing constraints.
In particular, for $\widetilde{M}=100$~TeV shown in the lower panel of \cref{fig:loophole}, cosmological constraints from BBN, the CMB, and $\gamma$-ray searches vanish below the equilibrium line (\textcolor{black}{black}) throughout the mesoscopic and microscopic regions of parameter space.

The bounds from $\dNeff$ at BBN can still exclude portions of parameter space that evade all other cosmological constraints in this setup. If the SM and dark gluons thermalize, i.e. above the \textcolor{black}{black} line in \cref{fig:loophole}, then $\dNeff$ is determined solely by the relativistic degrees of freedom present at the time of decoupling and at the epoch of the $N_\mathrm{eff}$ measurement (e.g., BBN or the CMB), as described in \Eq{Neff_def}.
In this scenario, the new \textcolor{turquoise}{turquoise} region at larger $\Lambda$ originates from glueballs decaying early into the massless $\phi$ particles, which subsequently contribute to $\dNeff$. The \textcolor{turquoise}{turquoise} region at smaller $\Lambda$ corresponds to values where dark gluons themselves survive until BBN and act as additional relativistic species, reproducing the same qualitative behavior found in the original model without an invisible decay channel.

Crucially, for sufficiently small values of $\widetilde{M}$, there exists a band of $\Lambda$ for which the \textcolor{black}{black} line is no longer excluded by any cosmological observable.
Thus, in the presence of such an invisible decay channel for glueballs, discovering mesoscopic or microscopic quirks does not necessarily prove an upper bound on $\Trh$ of the universe. 
However, even with this additional decay mode, discovering macroscopic and megascopic quirks still continues to enforce stringent constraints on $\Trh$. 

Unlike the microscopic and the mesoscopic ranges, the $\dNeff$ bounds in the macroscopic quirk range cannot be avoided by introducing a new decay channel for glueballs. 
In this region, the constraints arise from the mere existence of dark gluons and their irreducible freeze-in abundance.

Our aim in this exercise has been to demonstrate that, although it is in principle possible to sever the link between a collider discovery of quirks and an upper bound on the reheating temperature, doing so requires nontrivial model-building even in the simplest extensions (such as introducing an invisible glueball decay channel). 
Even then, we have not been able to break the link completely -- only for \textit{microscopic} and \textit{mesoscopic} quirks.
The very need for such additional structure suggests that any scenario evading this connection would itself point to richer BSM dynamics beyond the quirks discovered at colliders.
Either way, this provides strong motivation for dedicated collider searches for quirks.

\end{spacing}

\bibliography{ref}
\bibliographystyle{utphys}

\end{document}